\documentstyle[12pt]{article}

\def\fnote#1#2{\begingroup\def\thefootnote{#1}\footnote{#2}
    \addtocounter{footnote}{-1}\endgroup}

\def\email{makoto.natsuume@kek.jp}
\def\okamura{okamura@skyrose.phys.ocha.ac.jp}
\def\sato{sato@th.phys.titech.ac.jp; JSPS Research Fellow}

\newcommand{\be}{\begin{equation}}
\newcommand{\ee}{\end{equation}}
\newcommand{\bea}{\begin{eqnarray}}
\newcommand{\eea}{\end{eqnarray}}

\newcommand{\expv}[1]{\mbox{$\langle #1 \rangle$}}
\newcommand{\eq}[1]{(\ref{eq:#1})}

\newcommand{\rp}{r_+}

\newcommand{\half}{{1 \over 2}}

\def\rslash{\partial\kern-0.026em\raise0.17ex\llap{/}%
\kern0.026em\relax}
\def\Dslash{D\kern-0.15em\raise0.17ex\llap{/}\kern0.15em\relax}
\def\sqr#1#2{{\vcenter{\hrule height.#2pt
      \hbox{\vrule width.#2pt height#1pt \kern#1pt
          \vrule width.#2pt}
      \hrule height.#2pt}}}

\newcommand{\cH}{{\cal H}}
\newcommand{\cJ}{{\cal J}}
\newcommand{\cL}{{\cal L}}
\newcommand{\cM}{{\cal M}}
\newcommand{\cV}{{\cal V}}
\newcommand{\hH}{{\hat H}}
\newcommand{\back}[1]{ \stackrel{\circ}{#1} }
\newcommand{\og}{ {\back{g}} }
\newcommand{\oh}{ {\back{h}} }
\newcommand{\oD}{ {\back{D}} }
\newcommand{\oG}{ {\back{G}} }
\newcommand{\oS}{ {\back{S}} }

\newcommand{\opi}{ {\back{\pi}} }

\newcommand{\speta}{ {{}^{(3)}\!\eta} }

\begin{document}

\pagestyle{empty} 

\begin{flushright}
\end{flushright}

\begin{flushright}
	KEK-TH-652, OCHA-PP-144, TIT/HEP-432\\
	hep-th/9910105\\
\end{flushright}

\begin{center}
{\large \bf Three-Dimensional Gravity with Conformal Scalar} \\ \vspace{4pt} 
{\large \bf and Asymptotic Virasoro Algebra}

\vspace{16pt}
Makoto Natsuume, $^{1}$ \fnote{*}{\email} Takashi Okamura, $^{2}$ \fnote{\dag}{\okamura} and Masamichi Sato $^{3}$ \fnote{\ddag}{\sato}

\vspace{16pt}
{\sl $^{1}$ Theory Division\\
Institute of Particle and Nuclear Studies\\
KEK, High Energy Accelerator Research Organization\\
Tsukuba, Ibaraki, 305-0801 Japan}

\vspace{8pt}

{\sl $^{2}$ Department of Physics\\
Ochanomizu University\\
1-1 Otsuka, 2 Bunkyo-ku, Tokyo, 112-8610 Japan}

\vspace{8pt}

{\sl $^{3}$ Department of Physics\\
Tokyo Institute of Technology \\
Oh-okayama, Meguro, Tokyo, 152-8551 Japan}

\vspace{12pt}
{\bf ABSTRACT}

\end{center}

\begin{minipage}{4.8in}
Strominger has derived the Bekenstein-Hawking entropy of the BTZ black hole using asymptotic Virasoro algebra. We apply Strominger's method to a black hole solution found by Mart\'{\i}nez and Zanelli (MZ). This is a solution of three-dimensional gravity with a conformal scalar field. The solution is not $AdS_3$, but it is asymptotically $AdS_3$; therefore, it has the asymptotic Virasoro algebra. We compute the central charge for the theory and compares Cardy's formula with the Bekenstein-Hawking entropy. It turns out that the functional form does agree, but the overall numerical coefficient does not. This is because this approach gives the ``maximum possible entropy" for the numerical coefficient.
\end{minipage}


\vfill
\pagebreak

\pagestyle{plain}	
\setcounter{page}{1}	

\baselineskip=16pt

\section{Introduction}

Using D-brane technology \cite{SV}, one now believes that the Bekenstein-Hawking entropy of a black hole is a true statistical entropy. There remain many issues however.

For example, the D-brane approach depends on the details of D-brane dynamics and is complicated. However, black hole thermodynamics is a universal feature of metric theories of gravity \cite{wald}. Moreover, the Bekenstein-Hawking entropy was originally derived using quantum field theory on curved spacetime. Therefore, any microscopic theory should reproduce the Bekenstein-Hawking entropy if its low-energy action is written in terms of metric. A derivation of the black hole entropy is just a consistency check of a microscopic theory. 

Thus, the details of a microscopic theory should not be relevant in order to understand the entropy. One would like to know the necessary and sufficient ingredients to understand the entropy. Strominger's work is an important step in this respect \cite{andy}.

Three-dimensional anti-deSitter space ($AdS_3$) has the asymptotic symmetry group which is generated by two copies of Virasoro algebra \cite{BH}. The central charge of the algebra is given by
\be
c = \frac{3l}{2G},
\label{eq:cc}
\ee
where $G$ is the three-dimensional Newton's constant and $\Lambda=-1/l^2$ is the cosmological constant. Then, if one considers quantum gravity on $AdS_3$, the physical states at infinity must form a representation of this algebra. The asymptotic density of states of a conformal field theory (CFT) is given by Cardy's formula:
\be
S_{asymp} = 2 \pi \sqrt{\frac{c \Delta}{6}} + 2 \pi \sqrt{\frac{c \bar{\Delta}}{6}},
\label{eq:cardy}
\ee
where $\Delta$ and $\bar{\Delta}$ are the eigenvalues of $L_0$ and $\bar{L}_0$. Applying this formula to the Ba\~{n}ados-Teitelboim-Zanelli (BTZ) black hole \cite{BTZ}, Strominger has obtained the Bekenstein-Hawking entropy for the black hole with a precise numerical coefficient. For the BTZ black hole, $\Delta$ and $\bar{\Delta}$ are given by the mass and angular momentum as follows:
\bea
\Delta	&=& \frac{1}{2} (lM+J), \\
\bar{\Delta}	&=& \frac{1}{2} (lM-J).
\eea

The derivation is very powerful and does not depend on the details of the microscopic theory. However, there are many problems as well \cite{carlip-9806}. One of the most important problems is the relevance of the asymptotic geometry and where the degrees of freedom actually live. 

Strominger's argument depends on the asymptotic symmetries at spatial infinity. On the contrary, the Bekenstein-Hawking entropy depends only on the area of a black hole horizon. One usually regards this as an indication that the degrees of freedom relevant for the entropy live on the horizon. If so, what is really important is the near-horizon geometry and not the asymptotic geometry. In fact, Strominger's argument has been applied to various higher dimensional black holes whose near-horizon geometries are the BTZ black hole and has reproduced the correct entropy \cite{nearHorizon}. There are several possibilities why it works:

\begin{enumerate}

\item In light of $AdS$/CFT correspondence \cite{maldacena,witten,GKP}, there are two possibilities:

\begin{enumerate}
\item The branes are indeed located there. This could be possible since asymptotic infinity is not really infinity; one only looks at the near-horizon geometry.
\item Martinec points out another possibility \cite{martinec-9804}; the entropy comes from brane dynamics at the horizon, but the conformal anomaly of the branes is transported at spatial infinity by an ``anomaly inflow mechanism."
\end{enumerate}

\item From the viewpoint of pure gravity, this may be because of the trivial nature of three-dimensional gravity which has no bulk degrees of freedom. For instance, three-dimensional gravity can be written as a boundary Liouville theory \cite{CHD}.

\end{enumerate}
Of course, the Liouville theory has too few degrees of freedom to account for the entropy; its effective central charge is $c_{eff}=1$. But we regard this as an indication that the Liouville theory is just the master field, not the microscopic description \cite{carlip-9806,martinec-9804,HKL}.
 
Even though three-dimensional pure gravity has no bulk degrees of freedom, the entropy could be non-zero. One should distinguish the low-energy degrees of freedom discussed here and the microscopic degrees of freedom which constitute the Bekenstein-Hawking entropy. The Bekenstein-Hawking entropy predicts that any sensible quantum gravity must have the degrees of freedom which are necessary to account for the entropy. How such degrees of freedom are actually realized may differ for each quantum theory. For example, it may come from the massive stringy degrees of freedom \cite{GKS}. Or it may come from the twisted sectors in pure gravity \cite{banados-99}. It would be interesting to find how such degrees of freedom actually explain the central charge \eq{cc}, but as far as Strominger's approach concerns, it is not really necessary. 

In this paper, we consider three-dimensional gravity with a conformal scalar field. We apply Strominger's argument to a black hole solution of the theory, the Mart\'{\i}nez and Zanelli (MZ) solution \cite{MZ}. The solution is not $AdS_3$ but it is asymptotically $AdS_3$ (See Section~\ref{sec-AAdS} for the definition of ``asymptotically $AdS_3$"); therefore, it has the asymptotic Virasoro algebra. If Strominger's method works due to the triviality of three-dimensional gravity, it will not work for such a theory. Thus, our approach is somewhat different from Strominger's one. 

In Section~\ref{sec-mz}, we review the MZ solution and discuss the properties of the solution. In Section~\ref{sec-hamiltonian}, we discuss the Hamiltonian formalism and calculate the central charge \'{a} la Brown-Henneaux. In Section~\ref{sec-discussion}, using these results, we compare Cardy's formula with the Bekenstein-Hawking entropy. We found that the functional form agrees with the boundary CFT prediction. The overall numerical coefficient does not agree however; this is because this approach gives the ``maximum possible entropy" for the numerical coefficient \cite{carlip-9906}. In this sense, the discrepancy in the numerical coefficient may indicate that this method does not work perfectly when there are bulk low-energy degrees of freedom and may support Possibility 2. 
In Appendix~\ref{sec-limit}, we summarize a number of different expansions considered in this paper. In Appendix~\ref{sec-ads/cft}, we discuss $AdS$/CFT correspondence to obtain the central charge. This is the first attempt to study Strominger's method applied to asymptotically $AdS_3$ black holes other than the BTZ black hole.

\section{The Mart\'{\i}nez-Zanelli Black Hole}\label{sec-mz}

If Ricci scalar is constant, a metric satisfies the Brown-Henneaux's boundary conditions \cite{BH}. Solving the constant Ricci scalar for a three-dimensional static metric, one can easily see that the metric satisfies the boundary conditions. Now, since the field equation can be written as
\be
R_{\mu \nu} - \frac{1}{2} g_{\mu \nu} R - \frac{1}{l^2} g_{\mu \nu} = 8 \pi G T_{\mu \nu},
	\label{eq:metricEOM}
\ee
the Ricci scalar is in fact constant if the matter stress tensor $T_{\mu \nu}$ is traceless:
\be
R = -\frac{6}{l^2}.
\ee
This is a reason why we consider the conformal scalar; the conformal scalar has a traceless stress tensor.

We consider the action given by
\be
S = \int_M d^{3}x \sqrt{-g} \left\{ \frac{R + 2l^{-2}}{16 \pi G} - \frac{1}{2} (\nabla \phi)^2 - \frac{\xi}{2} R \phi^2 \right\} + B',
\label{eq:action}
\ee
where $\xi=\frac{d-2}{4(d-1)}=\frac{1}{8}$. The surface term $B'$ should be included so as to eliminate the second derivatives of the metric. The field equations are given by Eq.~\eq{metricEOM} and the matter field equation,
\be
\nabla^2 \phi - \xi R \phi = 0.
	\label{eq:matterEOM}
\ee
The matter stress tensor $T_{\mu \nu}$ is given by
\be
T_{\mu \nu} = \nabla_{\mu} \phi \nabla_{\nu} \phi - \frac{1}{2} g_{\mu \nu} (\nabla \phi)^2 
	+ \xi \{ g_{\mu \nu} \nabla^2 - \nabla_{\mu} \nabla_{\nu} + R_{\mu \nu} - \frac{1}{2} g_{\mu \nu} R \} \phi^2.
\label{eq:em}
\ee
One can easily check that the matter stress tensor \eq{em} is traceless.

The MZ solution is 
\bea
ds^2 & = & - \frac{1}{l^2} \left(r + \frac{r_+}{2}\right)^2 \left(1 - \frac{r_+}{r}\right) dt^2 
	+ \frac{l^2 dr^2}{(r + \frac{r_+}{2})^2(1 - \frac{r_+}{r})} + r^2 d \varphi^2,
		\label{eq:mz} \\
\phi & = & \sqrt{\frac{r_+}{\pi G (2r + r_+)}},
\eea
where $r=r_+$ is the horizon which is related to the mass by
\be
M=\frac{3}{32 G} \left(\frac{r_+}{l}\right)^2.
\ee

We make a couple of comments on the solution. First, the metric \eq{mz} approaches $AdS_3$ asymptotically. However, the solution is not $AdS_3$ in general. To see this, note that $AdS_3$ is conformally flat. The standard Weyl tensor vanishes identically in three dimensions, so it is not suitable for our purpose. But there is a conformally invariant tensor which plays a role analogous to that of the Weyl tensor \cite{eisenhart}:
\be
C_{\lambda\mu\nu} = 
	\nabla_{\nu} R_{\lambda \mu} - \nabla_{\mu} R_{\lambda \nu} 
	-\frac{1}{4} (g_{\lambda \mu} \partial_{\nu} - g_{\lambda \nu} \partial_{\mu}) R.
\ee
A three-geometry is conformally flat if and only if $C_{\lambda\mu\nu}=0$. For the metric \eq{mz}, the tensor $C_{\lambda\mu\nu}$ vanishes only asymptotically. 

Second, since we have a constant Ricci scalar, the conformal scalar $\phi$ has $m^2 = -\frac{3}{4l^2}$ from Eq.~\eq{matterEOM}; the scalar is a tachyon. This is not a problem. The stability on $AdS_d$ only requires that $m^2 \geq -\frac{(d-1)^2}{4l^2}$ \cite{stability}. Conformal scalars in any dimensions satisfy the bound since $m^2 = -\frac{d(d-2)}{4l^2}$. However, the MZ black hole is not stable under linear perturbations of the metric \cite{MZstability}. This could cause a problem if one wants to discuss its thermodynamics. This is not really our purpose however; for our purpose, it is sufficient if the entropy makes sense.

Third, the solution has a non-trivial scalar field; moreover, the scalar field is regular everywhere. On the other hand, the no-hair theorems require that scalar fields which are regular be vanish \cite{hair}. This is partly due to the fact that the spacetime in question is not asymptotically flat. 

Bekenstein's proof is for four-dimensional spacetime without a cosmological constant. However, the setup, itself, can be used for any dimensions and for the cases with a cosmological constant. His proof uses the fact that the volume integral of a positive definite function which is made from scalar fields is equal to surface integrals at the horizon and at asymptotic infinity. The surface integral at infinity vanishes for asymptotically flat solutions. The surface integral at the horizon vanishes for the scalar fields bounded on the horizon. Therefore, the volume integral has to vanish and the only way is for the scalar fields to vanish identically. Thus, any non-trivial scalar field has to diverge at the horizon. However, the surface integral at asymptotic infinity does not vanish for the MZ solution; therefore, the scalar field does not have to vanish. 

Now, the entropy of the black hole is given by
\be
S_{BH} = \frac{\pi r_+}{3G}
\label{eq:entropy1}
\ee
or
\be
S_{BH} = \frac{4\pi}{3} l \sqrt{\frac{2M}{3G}}
\label{eq:entropy2}
\ee
in terms of the mass. The entropy does not satisfy the area law $S_{BH}=A/(4G)$. This is because the action \eq{action} is not in the Einstein-Hilbert action form. Rather, the action \eq{action} is written as 
\be
S = \frac{1}{16 \pi G} \int_M d^{3}x \sqrt{-g} (1 - \pi G \phi^2) R + \cdots.
\ee
One can understand that the discrepancy arises because the Newton's constant at the horizon is scaled by the factor $(1-\pi G \phi^2)$. By a conformal transformation to the Einstein metric,
\be
g_{\mu\nu}^{E} = (1-\pi G \phi^2)^2 g_{\mu\nu},
\ee
one can check the area law is satisfied in the Einstein metric.

The mass and the Hawking temperature of a black hole are physical quantities. These quantities do not change under a conformal transformation \cite{JK}; the conformal transformation is just a change of variables. So, the black hole entropy is a physical quantity as well by the first law of black hole thermodynamics. However, the horizon area does change under the transformation. Therefore, the area law is not satisfied in all frames; it is satisfied only in the Einstein metric. The area law in a general frame is given by Wald's formula \cite{wald}:
\be
S_{BH} = -2\pi \int_{\sigma} \frac{\delta {\cal L}}{\delta R_{\mu\nu\rho\sigma}} n_{\mu\nu} n_{\rho\sigma},
\ee
where ${\cal L}$ is a Lagrangian, $n_{\mu\nu}$ is the binormal to the horizon $\sigma$ with the normalization $n^2=-2$, and the functional derivative is taken by formally regarding the Riemann tensor as a field which is independent of the metric.
\footnote{Strictly speaking, $S_{BH}$ defined by Wald's formula has not been proven to satisfy the second law of black hole thermodynamics \cite{wald}.}
Applying the formula to the action \eq{action}, we get
\be
S_{BH} = \frac{(1-\pi G \phi^{2}_{+}) A}{4G},
\ee
where $\phi_{+}$ is the value of the conformal scalar at the horizon. This expression agrees with Eq.~\eq{entropy1}.

\section{Hamiltonian formulation}\label{sec-hamiltonian}

In this section, we summarize the asymptotic symmetry and the Hamiltonian formulation \cite{RT} of three-dimensional Einstein gravity with a conformal scalar field. We then compute the central charge of the theory.

\subsection{Asymptotically Anti-deSitter Space}\label{sec-AAdS}

We are interested in any black hole solutions which are $AdS_3$ asymptotically, so we first define ``asymptotically $AdS_3$":
\begin{enumerate}
  \item[(i)]  They should contain the MZ black hole solution.
  \item[(ii)]  They should be invariant under the $AdS_3$ group $O(2,2)$
         at spatial infinity.
  \item[(iii)]  They should make the surface integrals
         associated with the generators of $O(2,2)$ finite.
\end{enumerate}
These are the same conditions as imposed by Henneaux and Teitelboim for asymptotically $AdS_4$
\cite{H-T85}.

We henceforth use the zero-mass black hole as the reference spacetime. Although the reference spacetime is the source of a number of issues (see Section~\ref{sec-hws}), we adopt it for the time being. In order to explicitly represent the asymptotically $AdS_3$ conditions, consider the coordinate system in which the zero-mass black hole reads
\begin{equation}
	ds^2=\og_{\mu\nu} dx^\mu dx^\nu
	= -{r^2 \over l^2} dt^2 + {l^2 \over r^2} dr^2 + r^2 d\varphi^2.
\label{eq:zeromass}
\end{equation}
Then, the components of the $AdS_3$ Killing vector $\eta$ behave as
\renewcommand{\arraystretch}{1.5}
\be
\begin{array}{lll}
        \speta{}^t = O(1), & \speta{}^r = O(r), & \speta{}^\varphi=O(1),
\nonumber \\
        \partial_r \speta{}^t = O(r^{-3}), & \partial_r \speta{}^r = O(1), &
\partial_r \speta{}^\varphi=O(r^{-3}),
\end{array}
\label{eq:BCforta}
\ee
\renewcommand{\arraystretch}{1}%
where $\speta{}^\alpha$ ($\alpha=t,r,\varphi$) are the components of the vector $\eta$ in the spacetime coordinate $\eta= \speta{}^\alpha~\partial_\alpha$. We will also use $\eta^\mu$ ($\mu=\perp,r,\varphi$) for the components of the same vector $\eta$, which is decomposed into $\eta= \eta^\perp n+\eta^i \partial_i$ ($i=r,\varphi$) where $n$ is a unit normal vector to the time slice. The components $\eta^\perp$ and $\eta^i$ describe the normal and tangential components of a hypersurface deformation. Using the lapse $N^\perp$ and the shifts $N^i$, they are related to the spacetime components by
\begin{eqnarray}
	& &\eta^\perp = N^\perp \speta{}^t,
\\
	& &\eta^i = \speta{}^i + {N^i \over N^\perp} \eta^\perp.
\label{eq:relsptodef}
\end{eqnarray}

{}From the conditions~(i) and (ii), we are led to the boundary conditions for the metric perturbation $q_{\mu\nu}=g_{\mu\nu}-\og_{\mu\nu}$,
\begin{equation}
        q_{ab}=O(1), \hspace{1cm} q_{ra}=O(r^{-3}),
        \hspace{1cm} q_{rr}=O(r^{-4}),
\label{eq:qasympt}
\end{equation}
where $a, b=t, \varphi$ and
\begin{equation}
        \phi=O(r^{-1/2}).
\label{eq:phiasympt}
\end{equation}
The asymptotic behavior of $\phi$ is motivated by the MZ solution and is kept by the coordinate transformation of the $AdS_3$ group.

By using Eq.(\ref{eq:qasympt}), the asymptotic behavior of the $AdS_3$ Killing vector ({\ref{eq:BCforta}) is rewritten in terms of $\eta^\mu$ by
\renewcommand{\arraystretch}{1.5}
\be
\begin{array}{lll}
        \eta^\perp = O(r), & \eta^r = O(r), & \eta^\varphi=O(1),
\nonumber \\
        \partial_r \eta^\perp = O(1), & \partial_r \eta^r = O(1), &
\partial_r \eta^\varphi=O(r^{-3}).
\end{array}
\label{eq:BCfortaII}
\ee
\renewcommand{\arraystretch}{1}%

As Brown and Henneaux showed for pure gravity
\cite{BH},
the asymptotic symmetry which preserves the asymptotic conditions (\ref{eq:qasympt}) and (\ref{eq:phiasympt}) is actually extended into the pseudo-conformal group in two dimensions:
\begin{eqnarray}
	\speta{}^t &=& l \left[ T^+(x^+) + T^-(x^-)
	+{l^2 \over 2 r^2} ( \partial_+^2 T^+ + \partial_-^2 T^- ) \right]
	+ O(r^{-4}),
\nonumber \\
	\speta{}^r &=& - r~ ( \partial_+ T^+ + \partial_- T^- ) + O(r^{-1}),
\label{eq:virasorocomp} \\
	\speta{}^\varphi &=& T^+ - T^-
	-{l^2 \over 2 r^2} ( \partial_+^2 T^+ - \partial_-^2 T^- )
	+ O(r^{-4}),
\nonumber
\end{eqnarray}
where
\begin{equation}
	x^\pm = {t \over l} \pm \varphi.
\end{equation}
One can indeed check that the above diffeomorphisms obey the Virasoro algebra. Denoting the diffeomorphisms with $T^\pm_i~(i=1,2,3)$ as $\eta_i$, one finds
\begin{equation}
	T^\pm_3 = 2 \left( T^\pm_1 \partial_\pm T^\pm_2
	- T^\pm_2 \partial_\pm T^\pm_1 \right),
\end{equation}
where $\eta_3 = [ \eta_1, \eta_2]$. Therefore, $L_n$ and ${\bar L}_n$ ($-\infty < n < \infty$) which generate the diffeomorphisms with $T^+=e^{i n x^+}/2$ and $T^-=e^{i n x^-}/2$ respectively, obey the algebra
\begin{eqnarray}
	i [ L_n,~L_m ] &=& (n-m) L_{n+m},
\\
	i [ {\bar L}_n,~{\bar L}_m ] &=& (n-m) {\bar L}_{n+m},
\\
	{} [ L_n,~{\bar L}_m ] &=& 0.
\end{eqnarray}

Since by definition
\begin{eqnarray}
	\partial_t &=& { L_0 + {\bar L}_0 \over l},
\\
	\partial_\varphi &=&  L_0 - {\bar L}_0,
\end{eqnarray}
the mass and angular momentum of a black hole are related to the charges associated with $L_0$ and ${\bar L}_0$ as
\begin{eqnarray}
	M &=& { L_0 + {\bar L}_0 \over l},
\\
	J &=&  L_0 - {\bar L}_0.
\end{eqnarray}

\subsection{Canonical Realization of Asymptotic Symmetry}

Our task now is to calculate the central charge in the canonical realization of the algebra.

Using the standard ADM decomposition, we obtain the bulk Hamiltonian
\begin{equation}
        H[ N ] = \int d^2 x \left[ N^\perp \cH_\perp + N^i \cH_i \right],
\label{eq:bulkHam}
\end{equation}
where $N^\perp$ and $N^i$ are the lapse and the shift functions respectively. The Hamiltonian and momentum constraints are given by
\begin{eqnarray}
	& &\cH_\perp = {1 \over 2 \sqrt{h}} \pmatrix{ \pi^{ij} & p \cr}
	\cM^{-1} \pmatrix{ \pi^{kl} \cr p \cr} + \sqrt{h}~ \cV,
\label{eq:bulkHamconst} \\
	& &\cH_i = -2 \sqrt{h} D_j \left(
	{\pi_i{}^j \over \sqrt{h}} \right) + p D_i \phi,
\label{eq:bulkmomconst}
\end{eqnarray}
where $\pi^{ij}$ and $p$ are the canonical conjugate momenta of $h_{ij}$ and $\phi$ respectively, and
\begin{eqnarray}
	& &\cM^{-1} = \pmatrix{ {32 \pi G \over U} G_{ijkl}
	+ { 4 U'^2 \over U} h_{ij} h_{kl} & -2 U' h_{ij} \cr
	-2 U' h_{kl} & U \cr },
\\
	& &G_{ijkl} = h_{i(k}h_{l)j}- h_{ij} h_{kl},
\hspace{1cm} U = 1 - {\xi \over 2} (16\pi G) \phi^2,
\\
	& &\cV = -{1 \over 16\pi G} \left( U {}^{(2)}\!R + 2 l^{-2} \right)
	+\half \left( D\phi \right)^2 - \xi \Delta \phi^2.
\end{eqnarray}
Here, $U'=\frac{\partial U}{\partial \phi}$ and ${}^{(2)}\!R$ is the two-dimensional Ricci scalar.

Since we consider spacetimes with open spacelike surfaces, we must pay attention to boundary terms. The boundary terms are necessary in order to make the functional derivative of the Hamiltonian well-defined. From Eqs.(\ref{eq:qasympt}) and (\ref{eq:phiasympt}), we can read off the boundary conditions for the canonical variables as
\renewcommand{\arraystretch}{1.5}
\be
\begin{array}{lll}
	q_{rr}=O(r^{-4}), & q_{r\varphi}=O(r^{-3}), &
q_{\varphi\varphi}=O(1), \\
	\pi^{rr}=O(r^{-1}), & \pi^{r\varphi}=O(r^{-2}), &
\pi^{\varphi\varphi}=O(r^{-5}), \\
	\phi=O(r^{-1/2}), & p=O(r^{-3/2}). &
\end{array}
\label{eq:BCforcv}
\ee
\renewcommand{\arraystretch}{1}%
Under the asymptotic behavior of the canonical variables \eq{BCforcv} and the surface deformation vector $\eta$ \eq{BCfortaII}, the boundary terms become
\begin{eqnarray}
	\delta J[\eta] &=& \delta \cJ[\eta] - \xi \oint d\oS_i
	(\eta^\perp \phi^2)^2
	D^i \left\{ {\delta (\phi^{-2}) \over \eta^\perp} \right\},
\label{eq:deltaJ} \\
	\cJ[\eta] &=& \oint d\oS_i \left[
	2~ { \left( \eta_j \pi^{ij} \right) \over \sqrt{\oh} }
	 + {1 \over 16\pi G} \oG{}^{ijkl} \left\{ \eta^\perp \oD_j q_{kl}
	- q_{kl} (\oD_j \eta^\perp) \right\} \right],
\label{eq:JfromGrav}
\end{eqnarray}
where $\delta$ is any variation in the configuration space of asymptotically $AdS_3$. Note $\cJ[\eta^\mu]$ is the surface term which arises for pure gravity.

In general, we may not be able to write the second term of Eq.(\ref{eq:deltaJ}) in terms of a total derivative. However, the MZ solution has ``no-scalar hair," which means that the second term in Eq.(\ref{eq:deltaJ}) vanishes. Thus, we impose the ``no-scalar hair" condition:
\begin{eqnarray}
	& &\delta B(t,~\varphi)=0,
\label{eq:BCforfinitecharge} \\
	& &{1 \over \phi^2} = r~ A(t,~\varphi) + B(t,~\varphi) + O(r^{-1}).
\label{eq:defB}
\end{eqnarray}
We must check whether the ``no-scalar hair" condition is preserved under the action of the conformal group (\ref{eq:virasorocomp}). Suppose $\delta B =0$ in some coordinate system satisfying Eqs.(\ref{eq:qasympt}) and (\ref{eq:phiasympt}). Using the transformation (\ref{eq:virasorocomp}), we obtain
\begin{eqnarray}
	\delta_{Lie} (\phi^{-2}) &=& - \cL_\eta \phi^{-2},
\nonumber \\
	&=& 2 r \left( T^+ \partial_+ + T^- \partial_-
	- { \partial_+ T^+ + \partial_- T^- \over 2} \right) A(t, \varphi)
\nonumber \\
	& &+ 2 \left( T^+ \partial_+ + T^- \partial_- \right) B(t, \varphi)
	+ O(r^{-1}).
\end{eqnarray}
Thus the transformed $\phi$ also satisfies the ``no-scalar hair" condition. Therefore, the Hamiltonian $\hH$ with the well-defined functional derivative is
\be
\hH[\eta] = H[\eta] + \cJ[\eta]
\ee
under the asymptotically $AdS_3$ conditions and the ``no-scalar hair" condition. Note that we need only the leading order terms of $\eta$ in $1/r$ to evaluate the charges $\cJ[\eta]$.

The mass and angular momentum of a black hole are given by
\begin{eqnarray}
	& &M = \cJ[ \partial_t ] = { \cJ[ L_0 ] + \cJ[ {\bar L}_0 ] \over l},
\\
	& &J = \cJ[ \partial_\varphi ] = \cJ[ L_0 ] - \cJ[ {\bar L}_0 ].
\end{eqnarray}
Because we use the zero-mass black hole as the reference geometry, $M=-1/8G$ for the globally $AdS_3$.

We obtain the same surface term as the one for pure gravity; thus the central charge coincides with the pure gravity one by applying the same argument as Brown and Henneaux. We present their argument for completeness. Using the Dirac bracket $\{~,~\}_{DB}$ for the constraints $\cH_\mu \approx 0$, the algebra of the asymptotic symmetry becomes
\begin{equation}
	\left\{ \cJ[\eta_1], ~\cJ[\eta_2] \right\}_{DB} =
	\cJ[~ [\eta_1, \eta_2]~ ] + K[ \eta_1,~\eta_2 ],
\label{eq:algebracJ}
\end{equation}
where $K[ \eta_1,~\eta_2 ]$ is the central charge. The left-hand side is just the change in the charge $\cJ[\eta_1]$ under the surface deformation generated by $\cJ[\eta_2]$, that is,
\begin{equation}
	 \{ \cJ[ \eta_1 ],~\cJ[\eta_2] \}_{DB} = \cL_{\eta_2} \cJ[ \eta_1 ]
	= - \delta_{Lie(\eta_2)} \cJ[ \eta_1 ].
\end{equation}
The central charge may be obtained from Eq.(\ref{eq:algebracJ}), which is most easily evaluated on the $t=0$ surface of the reference spacetime $\og_{\mu\nu}$. Because the charge has been chosen so that it vanishes for the reference spacetime, then $\cJ[~[\eta_1,\eta_2]~]=0$ and the charge $\cJ[\eta_1]$, before the surface is deformed, is also zero. Thus, substituting $q_{\mu\nu} = \delta_{Lie(\eta_2)} \og_{\mu\nu}$ into Eq.~\eq{JfromGrav}, we get
\begin{eqnarray}
	K[\eta_1,~\eta_2] &=& - \delta_{Lie(\eta_2)} \cJ[ \eta_1 ]
\nonumber \\
	 &=& \oint d\oS_i \left[
	2~ { \eta_{1 j} \left( \cL_{\eta_2} \opi{}^{ij} \right) \over \sqrt{\oh} }
	\right.
\\
	& & \left. 
	+ {1 \over 16\pi G} \oG{}^{ijkl} \left\{ \eta_1^\perp \oD_j 
	( \cL_{\eta_2} \og_{kl} ) 
	- ( \cL_{\eta_2} \og_{kl} ) (\oD_j \eta_1^\perp) \right\} 
	\right] 
\nonumber \\
	& & - \left( \eta_1 \leftrightarrow \eta_2 \right).
\nonumber 
\label{eq:centralK}
\end{eqnarray}
Using Eq.(\ref{eq:virasorocomp}), we obtain the central charge for the Virasoro generators,
\begin{eqnarray}
	& &i K[ L_n,~L_m] = i K[ {\bar L}_n,~{\bar L}_m] =
	{ c \over 12} n^3~ \delta_{n+m,~0},
\\
	& &c={3l \over 2G},
\end{eqnarray}
that is,
\begin{equation}
	i~ \left\{ \cJ[ L_n ], ~\cJ[ L_m ] \right\}_{DB} =
	(n-m) \cJ[ L_{n+m} ] + {c \over 12} n^3~ \delta_{n+m,~0}.
\label{eq:Virasoroalgebra}
\end{equation}
The same holds for ${\bar L}_n$.
Thus we have the same central charge $c=3l/2G$ as pure gravity.

Note that the Virasoro algebra (\ref{eq:Virasoroalgebra}) has a non-standard form \cite{CH}. This is because the zero-mass black hole is used as the reference spacetime. If we take the globally $AdS_3$ as the reference spacetime, then $\cJ^{NS}[L_0]=\cJ[L_0]+c/24$, where $\cJ^{NS}$ is the new charge using the globally $AdS_3$ as the reference. The new charges $\cJ^{NS}[L_m]=\cJ[L_m]+ \delta_{n,~0}~c/24$ have the usual form of the Virasoro algebra:
\begin{equation}
	i~ \left\{ \cJ^{NS}[ L_n ], ~\cJ^{NS}[ L_m ] \right\}_{DB} =
	(n-m) \cJ^{NS}[ L_{n+m} ]
	+ {c \over 12} n(n^2-1)~ \delta_{n+m,~0}.
\label{eq:VirasoroalgebraII}
\end{equation}

\section{Discussion}\label{sec-discussion}

We found that the central charge and the $L_{0}$ eigenvalues of the black hole are unchanged from pure gravity results. Thus, if one simply applies Strominger's derivation, the asymptotic density of states estimated by the asymptotic symmetry group is the same as the pure gravity result:
\be
S_{asymp} = 4\pi \sqrt{\frac{c \Delta}{6}} = 2 \pi l \sqrt{\frac{M}{2G}}.
\label{eq:s_cft}
\ee
Thus, the functional form agrees with \eq{entropy2}, but the overall numerical coefficient does not. 

\subsection{Maximum Possible Entropy}

Note $S_{asymp} > S_{BH}$; this is because $S_{asymp}$ gives the ``maximum possible entropy" \cite{carlip-9906}. There is another solution in this theory, the BTZ black hole, which saturates the bound $S_{asymp} = S_{BH}$. Since Strominger's argument is insensitive to the details of the interior structure, the Virasoro algebra predicts the same answer for the BTZ black hole and for the MZ black hole. So, the CFT answer cannot be true for both solutions. The CFT prediction gives the numerical value of the larger entropy solution for a given mass. 

Then, one may be tempted to think that all we have done is just to rederive the BTZ black hole entropy in this theory, not the MZ black hole. However, if the CFT prediction simply gives the maximum possible entropy, the entropy of the MZ black hole does not have to have the same functional form as Cardy's formula. We believe that the fact that the functional form of the MZ black hole entropy agrees with the CFT prediction indicates that the CFT prediction gives the correct functional form even if a black hole entropy does not saturate the bound $S_{asymp} \geq S_{BH}$.

There still remains the question why the approach gives the maximum possible entropy. One possible answer is that we are counting the entropy from the matter field as well and somehow have to subtract them. The counting of matter entropy has been computed by various authors using the brick-wall model \cite{'tHooft:1985re} and it is consistent with the area formula in most cases. Thus, subtracting the matter entropy from $S_{asymp}$ changes only the numerical coefficient. However, such a computation generally depends on the regularization scheme.

\subsection{Lowest Virasoro Eigenvalues}\label{sec-hws}

Strominger's argument is impressive, but the derivation actually has several assumptions \cite{carlip-9806}. One important point is the lowest Virasoro eigenvalues; in order to use Cardy's formula, the lowest $L_0$ eigenvalues of the CFT should be $\Delta_0=\bar{\Delta}_0=0$. 

If the CFT is unitary, $\Delta_0^{NS} \geq 0$ so that there is no state with negative weights. For the BTZ black hole, the unitarity of the boundary CFT could be guaranteed by the underlying string theory with a Ramond-Ramond (R-R) background. On the other hand, the unitarity of our boundary CFT is not clear. 

This assumption is usually justified by regarding the BTZ black hole as excitations from the reference spacetime \cite{andy} [either the globally $AdS_3$ or the zero-mass black hole \eq{zeromass}]; because a reference geometry has zero mass and angular momentum by definition, it has $\Delta=\bar{\Delta}=0$. In this interpretation, the reference geometry is a highest weight state of the CFT. We assume that such a reference geometry has the lowest eigenvalues.

However, it is a different issue whether the zero-mass black hole \eq{zeromass} or the globally $AdS_3$ is the highest weight state which produces the MZ solution. In fact, it does not seem so. One can easily obtain the general Virasoro deformation using the asymptotic Killing vector \eq{virasorocomp}\cite{banados-99}. Because the Killing vector \eq{virasorocomp} and the geometry are the same as pure gravity, the most general deformation is the same as well. This includes the BTZ solution, but does not include the MZ solution since the scalar $\phi$ transforms as $\delta_{Lie} \phi = \speta{}^\alpha \partial_{\alpha} \phi = 0$. So, one can make only the BTZ black hole in this way. Thus, one may again conclude that we can only rederive the BTZ black hole entropy in this theory. 

This is not a problem. Cardy's formula sums over all states in the CFT, {\it i.e.}, the members of all conformal families. When there are many Verma modules, we do not know a priori which highest weight states we should choose. The zero-mass black hole does not have to be the highest weight state we are looking for; the derivation only requires that there is a highest weight state which produces the black hole in question. 

It is not clear whether there is a geometry which produces the MZ black hole by the Virasoro deformation. 
\footnote{One possible candidate is the zero-mass black hole metric \eq{zeromass} with $\phi=c/\sqrt{r}$, where $c$ is a constant. This is a solution of Eqs.~\eq{metricEOM} and \eq{matterEOM}.}
It is not clear either whether only one highest weight state produces the black hole. However, we do not know such a map between the classical deformations and the CFT description really makes sense. Anyway, the whole effect is an additive constant in the mass. So, we simply assume that there is a highest weight state which produces the MZ solution. If the CFT is unitary, all states lie in highest weight representations, so there should be such a state. 

There is another problem. Equation~\eq{s_cft} counts all states in the CFT with the same $L_0$ eigenvalues. Thus, one might suspect that we are counting the BTZ black hole entropy as well since both black holes have the same set of charges. We have no conclusive answer; we simply assume that these black holes somehow belong to the different sectors.

\subsection{Other Approaches}

In order to really understand the issue, it is desirable to have a microscopic description. Moreover, with such a description, one may be able to keep track of the conformal anomaly, {\it i.e.}, the anomaly induced on the ``effective string" on a brane intersection, the anomaly transported to infinity via anomaly inflow mechanism, and the anomaly of the boundary CFT. So, one can check whether the asymptotic CFT is really reflected on the horizon. Thus, it would be interesting to embed some asymptotically $AdS_3$ solutions in string theory (whose near-horizon geometry is not $AdS_3$). The solutions should necessarily be supersymmetric in order that the anomaly on the brane makes sense.

Even though we fail to reproduce the numerical coefficient of the Bekenstein-Hawking entropy, it may be possible to reproduce it using a different method. One possible approach is the Chern-Simons formalism \cite{banados-94}; the analysis has been claimed to be valid for a boundary located on any surface of constant $r$, in particular for the horizon. However, it is not clear how to choose the diffeomorphisms at the horizon. (Their diffeomorphisms reduce to Brown-Henneaux's asymptotic isometries at infinity.)

Another approach is Carlip's method, which is inspired by Strominger's analysis \cite{carlip-9906,carlip-9812}. He found a Virasoro algebra with a central charge which corresponds to the algebra of surface deformation of the $r-t$ plane that leaves the horizon fixed. The central charge is given by 
\be
c = \frac{3A}{2\pi G}.
\ee
The method applies to any black hole in any dimensions but applies only to pure gravity. It is interesting to see whether Wald's formula is reproduced when applied to gravity with various fields.

\vspace{.1in}
\begin{center}
    {\Large {\bf Acknowledgements} }
\end{center}
\vspace{.1in} 

We would like to thank V. Balasubramanian, R. G. Cai, A. Hosoya, N. Ishibashi, P. Kraus, J. Polchinski, and T. Sakai for useful discussions. The work of M.N. was supported in part by the Grant-in-Aid for Scientific Research (11740161) from the Ministry of Education, Science and Culture, Japan. The work of M.S. was supported in part by JSPS Research Fellowship for Young Scientists (10|0228).

\appendix

\section{The Classical Limit and Expansion Parameters}\label{sec-limit}

We consider a number of different expansions in this paper. In this section, we describe these expansions in detail. For simplicity, we consider black holes with $r_-=0$.

For the BTZ and the MZ black hole, the various length scales are related to physical quantities as follows (we neglect numerical coefficients):
\bea
c &\sim& \frac{l}{G}, \\
M &\sim& \frac{1}{G} \frac{\rp^2}{l^2}, \\
T &\sim& \frac{\rp}{l^2}, \\
S_{BH} &\sim& \frac{\rp}{G},
\eea

\begin{enumerate}

\item Semiclassical limit of pure gravity:
\label{semiclassical}

The semiclassical limit requires $c \gg 1$ \cite{andy} and thus
\be
l \gg G.
\ee

\item Cardy's formula:
\label{cardy}

Cardy's formula is an asymptotic formula valid at high energy. Thus one expects the formula is valid when
\be
L_0 = lM \gg c
\ee
and this implies
\be
\rp \gg l.
\ee
 
The condition is also necessary if we regard the globally $AdS_3$ as the state with the lowest Virasoro eigenvalues. Then, the black hole mass is given by $L_0^{NS}=lM+\frac{c}{24}$, where $M$ is the mass measured from the R-R ground state (the zero-mass black hole). Thus, Cardy's formula $S =4\pi\sqrt{c L_0^{NS}/6}$ is not the same as $S_{BH}$ due to the additive constant. In order to neglect the constant, one again must require $lM \gg c$.

One can also rewrite the condition as $T \gg 1/l$. This condition is actually too strong for D1/D5 system \cite{AGMOO} since this implies $N \gg Q_1 Q_5$. ($c=6Q_1 Q_5$ and $lM=N$) The effective string description of D1/D5 is valid even when $N \sim Q_1 Q_5$; this is the region where multiply wound strings dominate the entropy.

\item Backreaction of the emitted radiation:
\label{backreaction}

The standard treatment of black hole radiation neglects the backreaction of the radiation on the black hole. This is not possible if the emission changes the Hawking temperature by an amount comparable to the temperature \cite{PSSTW}. Thus, the heat capacity $C =|\partial E/\partial T|$ has to satisfy
\be
C \gg 1.
\ee
By the first law, this can be written as $|T \partial S/\partial T| \gg 1$; the entropy of the hole within the given thermal energy interval should be large. Thus,
\be
\rp \gg G
\ee

Conversely, the thermodynamics will only break down when the temperature is so low that $C \sim 1$. This happens at a temperature $T \sim G/l^2 \sim 1/(l c)$. In order for the black hole to be able to radiate at such low temperature, the mass gap for the theory should be of order $\delta M \sim G/l^2$ \cite{multiply}. This is a very long length scale; for D1/D5 system, this mass gap comes from multiply wound strings. 

\item Fluctuation of the geometry:
\label{geometry}

In order to neglect the fluctuation of the black hole geometry,
\be
\rp \gg r_c,
\ee
where $r_c=1/M$ is the Compton wavelength of the black hole. This implies
\be
\rp^3 \gg Gl^2.
\ee
If we recover $\hbar$, $G$ has a dimension of $(mass)^{-1}$. So the Planck mass should not be $1/G$ since this does not have a $\hbar$. It is impossible to form a mass scale from $G$ and $\hbar$ alone, so $(\hbar^2/(Gl^2))^{1/3}$ should be the Planck mass. If either $\rp \gg l$ or $\rp \gg G$, then this condition implies another. 

\end{enumerate}

In order for all expansions to be valid \cite{andy},
\be
\rp \gg l \gg G.
\ee
For the Schwarzschild black hole, the conditions \ref{backreaction} and \ref{geometry} both imply $\rp \gg \sqrt{G}$.

Since we have two basic length scales $l$ and $G$, one can form a variety of length scales \cite{reznik}. The length scales appeared in the above discussion are related by
\be
\frac{l^2}{G} \gg l \gg (l^2 G)^{1/3} \gg G.
\ee

\section{The Central Charge from Boundary Stress Tensor}\label{sec-ads/cft}

The central charge of $AdS_3$ found in the Hamiltonian formalism \cite{BH} has been reproduced by various authors. See refs.~\cite{banados-94} in the Chern-Simons formalism and refs.~\cite{HKL,BK,HS} in the framework of $AdS$/CFT correspondence \cite{maldacena,witten,GKP}. It would be interesting to obtain the central charge using the $AdS$/CFT correspondence. One possible approach is a boundary stress tensor proposal by Balasubramanian and Kraus \cite{BK}. 

Their starting point is the ``quasilocal stress tensor" proposal by Brown and York \cite{BY-93}. In order to obtain a definition of the mass, they proposed a tensor defined on the boundary of a given spacetime:
\be
T^{ab} = \frac{2}{\sqrt{-\gamma}} \frac{\delta S}{\delta \gamma_{ab}},
\label{eq:quasilocal}
\ee
where $\gamma_{ab}$ is the boundary metric and $a, b=t, \varphi$. The boundary metric is defined by writing the metric in an ADM-like decomposition:
\be
ds^2 = N^2 dr^2 + \gamma_{ab} dx^a dx^b.
\ee
The resulting stress tensor typically diverges. In order to obtain a finite stress tensor, they propose a subtraction by embedding a boundary with the same intrinsic geometry $\gamma_{ab}$ in some reference spacetime.

In light of $AdS$/CFT correspondence, Balasubramanian and Kraus have interpreted Eq.~\eq{quasilocal} as
\be
\expv{T^{ab}} = \frac{2}{\sqrt{-\gamma}} \frac{\delta S_{eff}}{\delta \gamma_{ab}},
\label{eq:stressdef}
\ee
where \expv{T^{ab}} is the expectation value of the CFT stress tensor. Then, the divergences which appear are simply the standard ultraviolet divergences of a quantum field theory and may be removed by adding local counterterms to the action which depend only on the boundary CFT \cite{witten}. On the other hand, the stress tensor of a two-dimensional CFT has a trace anomaly
\be
T^{(CFT)} = -\frac{c}{12} {}^{(2)}\!R,
\label{eq:anomaly}
\ee
where $c$ is the central charge and ${}^{(2)}\!R$ is the two-dimensional Ricci scalar. Thus, one can reproduce the central charge using this prescription.

However, in our case, we do not know the underlying boundary CFT so that it is not clear how to choose the counterterm action. In particular, we do not know what the conformal scalar corresponds to in CFT language.

One could go back to the original Brown-York prescription and try to find a suitable reference action and a reference spacetime. However, the prescription has drawbacks: it is not always possible to find an embedding and even if it is, such an embedding may not be unique. Moreover, it is not clear how to choose the reference action for gravity with scalar fields. As far as we are aware, there are three different prescriptions for the reference action \cite{CM-95,chan-96,CCM-96,chan-97,BL-98}:

\begin{enumerate}

\item CCM-CI \cite{CCM-96}: they have proposed the reference action from the requirement that mass is conformally invariant. However, the mass does not converge for the MZ solution if one chooses the zero-mass black hole as the reference spacetime. (We always choose the zero-mass black hole as the reference spacetime in this discussion.)

\item CCM-HH \cite{CCM-96}: this prescription is motivated by the Hawking-Horowitz (HH) proposal \cite{HH-95} to define mass. For the MZ solution, the mass is finite and is same as the one obtained in the Hamiltonian formalism. However, the mass is not conformally invariant. 

\item Bose-Lohiya \cite{BL-98}: this prescription is also motivated by the HH proposal, but mass is conformally invariant. For the MZ solution, the mass is finite, but does not agree with the Hamiltonian formalism.

\end{enumerate}

In conclusions, none of the prescriptions seem to be well-defined and can be used for our purposes. However, it would be interesting to study along this line further; this would give us a well-defined prescription of mass for gravity with scalar fields.



\begin{thebibliography}{999}


\bibitem{SV}
A. Strominger and C. Vafa, Phys. Lett. {\bf 379B} (1996) 99, {\tt hep-th/9601029}.

\bibitem{wald}
R. M. Wald, {\it Quantum field theory in curved space-time and black hole thermodynamics} (Chicago Univ. Press, Chicago, 1994) and references therein.

\bibitem{andy}
A. Strominger, JHEP {\bf 9802} (1998) 009, {\tt hep-th/9712251}.

\bibitem{BH}
J. D. Brown and M. Henneaux, Commun. Math. Phys. {\bf 104} (1986) 207.

\bibitem{BTZ}
M. Banados, C. Teitelboim, and J. Zanelli, Phys. Rev. Lett. {\bf 69} (1992) 1849, {\tt hep-th/9204099}; \\
M. Banados, M. Henneaux, C. Teitelboim, and J. Zanelli, Phys. Rev. {\bf D48} (1993) 1506, {\tt gr-qc/9302012}.

\bibitem{carlip-9806}
S. Carlip, Class. Quant. Grav. {\bf 15} (1998) 3609, {\tt hep-th/9806026}.

\bibitem{nearHorizon}
D. Birmingham, I. Sachs, and S. Sen, Phys. Lett. {\bf 424B} (1998) 275, {\tt hep-th/9801019};\\
V. Balasubramanian and F. Larsen, Nucl. Phys. {\bf B528} (1998) 229, {\tt hep-th/9802198};\\
E. Teo, Phys. Lett. {\bf B430} (1998) 57, {\tt hep-th/9803064};\\
M. Z. Iofa and L. A. Pando Zayas, Phys. Lett. {\bf 434B} (1998) 264, {\tt hep-th/9803083};\\
N. Kaloper, Phys. Lett. {\bf 434B} (1998) 285, {\tt hep-th/9804062};\\
G. Lopes Cardoso, Phys. Lett. {\bf B432} (1998) 65, {\tt hep-th/9804064};\\
M. Z. Iofa and L. A. Pando Zayas, Phys. Rev. {\bf D59} (1999) 064017, {\tt hep-th/9804129};\\
M. Cvetic and F. Larsen, Nucl. Phys. {\bf B531} (1998) 239, {\tt hep-th/9805097};\\
M. Cvetic and F. Larsen, Phys. Rev. Lett. {\bf 82} (1999) 484, {\tt hep-th/9805146};\\
F. Larsen, {\tt hep-th/9806071}.

\bibitem{maldacena}
J. M. Maldacena, Adv. Theor. Math. Phys. {\bf 2} (1998) 231, {\tt hep-th/9711200}.

\bibitem{witten}       
E. Witten, Adv. Theor. Math. Phys. {\bf 2} (1998) 253, {\tt hep-th/9802150}.

\bibitem{GKP}
S. S. Gubser, I. R. Klebanov, and A. M. Polyakov, Phys. Lett. {\bf 428B} (1998) 105, {\tt hep-th/9802109}.

\bibitem{martinec-9804}
E. J. Martinec, {\tt hep-th/9804111}.

\bibitem{CHD}
O. Coussaert, M. Henneaux, and P. van Driel, Class. Quant. Grav. {\bf 12} (1995) 2961, {\tt gr-qc/9506019}.

\bibitem{HKL}
S. Hyun, W. T. Kim, and J. Lee, Phys. Rev. {\bf D59} (1999) 084020, {\tt hep-th/9811005}. 

\bibitem{GKS}
A. Giveon, D. Kutasov, and N. Seiberg, Adv. Theor. Math. Phys. {\bf 2} (1998) 733, {\tt hep-th/9806194}.

\bibitem{banados-99}
M.~Banados,
Phys.\ Rev.\ {\bf D60} (1999) 104022,
{\tt hep-th/9903178}.

\bibitem{MZ}
C. Mart\'{\i}nez and J. Zanelli, Phys. Rev. {\bf D54} (1996) 3830, {\tt gr-qc/9604021}.

\bibitem{carlip-9906}
S.~Carlip,
Class.\ Quant.\ Grav.\ {\bf 16} (1999) 3327,
{\tt gr-qc/9906126}.


\bibitem{eisenhart}
L. P. Eisenhart, {\it Riemannian geometry} (Princeton Univ. Press, Princeton, 1926).

\bibitem{stability}
P. Breitenlohner and D. Z. Freedman, Phys. Lett. {\bf 115B} (1982) 197; Annals Phys. {\bf 144} (1982) 249;\\
L. Mezincescu and P. K. Townsend, Annals Phys. {\bf 160}(1985) 406.

\bibitem{MZstability}
C. Martinez, Phys. Rev. {\bf D58} (1998) 027501, {\tt gr-qc/9801091}.

\bibitem{hair}
J. E. Chase, Commun. Math. Phys. {\bf 19} (1970) 276; \\
J. D. Bekenstein, Phys. Rev. {\bf D5} (1972) 1239.

\bibitem{JK}
T. Jacobson and G. Kang, Class. Quant. Grav. {\bf 10} (1993) L201, {\tt gr-qc/9307002}.


\bibitem{RT}
T. Regge and C. Teitelboim, Annals Phys. {\bf 88} (1974) 286.

\bibitem{H-T85}
M. Henneaux and C. Teitelboim, Comm. Math. Phys. {\bf 98} (1985) 391.

\bibitem{Teitelboim73}
C. Teitelboim, Ann. of Phys. {\bf 79} (1973) 542.

\bibitem{CH}
O. Coussaert and M. Henneaux, Phys. Rev. Lett. {\bf 72} (1994) 183, {\tt hep-th/9310194}.



\bibitem{'tHooft:1985re}
G.~'t Hooft,
Nucl.\ Phys.\ {\bf B256} (1985) 727.

\bibitem{banados-94}
M. Banados, Phys. Rev. {\bf D52} (1996) 5816, {\tt hep-th/9405171};\\
M. Banados, T. Brotz, and M. Ortiz, Nucl. Phys. {\bf B545} (1999) 340, {\tt hep-th/9802076}.

\bibitem{carlip-9812}
S. Carlip, Phys. Rev. Lett. {\bf 82} (1999) 2828, {\tt hep-th/9812013}.


\bibitem{NS}
M. Natsuume and Y. Satoh, Int. J. Mod. Phys. {\bf A13} (1998) 1229, {\tt hep-th/9611041}.


\bibitem{AGMOO}
O. Aharony, S. S. Gubser, J. Maldacena, H. Ooguri, and Y. Oz, {\tt hep-th/9905111}.

\bibitem{PSSTW}
J. Preskill, P. Schwarz, A. Shapere, S. Trivedi, and F. Wilczek, Mod. Phys. Lett. {\bf A6} (1991) 2353.

\bibitem{multiply}
S. R. Das and S. D. Mathur, Phys. Lett. {\bf B375} (1996) 103, {\tt hep-th/9601152};\\
J. M. Maldacena and L. Susskind, Nucl. Phys. {\bf B475} (1996) 679, {\tt hep-th/9604042}.

\bibitem{reznik}
B. Reznik, Phys. Rev. {\bf D51} (1995) 1728, {\tt gr-qc/9403027}.


\bibitem{BK}
V. Balasubramanian and P. Kraus, {\tt hep-th/9902121}.

\bibitem{HS}
M. Henningson and K. Skenderis, JHEP {\bf 9807} (1998) 023, {\tt hep-th/9806087}.   

\bibitem{BY-93}
J. D. Brown and J. W. York, Jr., Phys. Rev. {\bf D47} (1993) 1407.

\bibitem{CM-95}
J. Creighton and R. B. Mann, Phys. Rev. {\bf D52} (1995) 4569, {\tt gr-qc/9505007}.

\bibitem{chan-96}
K. C. K. Chan, Phys. Rev. {\bf D55} (1997) 3564, {\tt gr-qc/9603038}.

\bibitem{CCM-96}
K. C. K. Chan, J. D. E. Creighton, and R. B. Mann, Phys. Rev. {\bf D54} (1996) 3892, {\tt gr-qc/9604055}.

\bibitem{chan-97}
K. C. K. Chan, {\tt gr-qc/9701029}.

\bibitem{BL-98}
S. Bose and D. Lohiya, Phys. Rev. {\bf D59} (1999) 044019, {\tt gr-qc/9810033}.

\bibitem{HH-95}
S. W. Hawking and G. T. Horowitz, Class. Quant. Grav. {\bf 13} (1996) 1487, {\tt gr-qc/9501014}.



\end{thebibliography}
\end{document}